\documentclass[oneside,english]{amsart}
\usepackage[T1]{fontenc}
\usepackage[latin9]{inputenc}
\usepackage{url}
\usepackage{amsthm}
\usepackage{setspace}
\makeatletter

\newcommand{\noun}[1]{\textsc{#1}}

\numberwithin{equation}{section}
\numberwithin{figure}{section}

\makeatother

\usepackage{babel}

\begin{document}

\title{Extending ArXiv.org to Achieve Open Peer Review and Publishing}

\author{Axel Boldt }

\address{Department of Mathematics\\
Metropolitan State University\\
Saint Paul, Minnesota, U.S.A.}

\email{Axel.Boldt@metrostate.edu}

\date{November 18, 2010}
\begin{abstract}
Today's peer review process for scientific articles is unnecessarily
opaque and offers few incentives to referees. Likewise, the publishing
process is unnecessarily inefficient and its results are only rarely
made freely available to the public. In this article we outline a
comparatively simple extension of\noun{ arXiv.org}, an online preprint
archive widely used in the mathematical and physical sciences, that
addresses both of these problems. Under the proposal, editors invite
referees to write public and signed reviews to be attached to the
posted preprints, and then elevate selected articles to {}``published''
status.
\end{abstract}
\maketitle

\section{The status quo}

In the system of peer review that is currently used in the sciences,
an editor invites one or more referees to review an article submitted
to a scientific journal. Based on the referees' recommendations, the
editor will accept the article, demand modifications, or reject it.
Referee reports are generally made available to the article's author
in anonymized form only and are not otherwise published. (Some journals
also anonymize the article to be refereed, even though ascertaining
the true author of a submission is usually a simple matter of using
an internet search engine.) 

The system as described is completely opaque to outside observers.
Neither the quality and timeliness of reviews, nor the standards of
a journal\textquoteright{}s editors, nor the extent of modifications
made after initial review, nor the number of times an article has
been rejected by other journals are publicly available. 

Other than professional integrity, referees have little legitimate
incentive to produce timely, fair and high-quality reviews. Since
the reviews are not published, referees are not accountable for their
work and cannot use it to bolster a case for professional advancement
or to improve their general standing in the academic community. Probably
the biggest (and most problematic) incentive for referees is the accumulation
of editor goodwill, to be expended during future article submissions.
It is also conceivable that some referees reject articles whose authors
they dislike or whose approach or results interfere with their own
research agenda. Finally, editors may circumvent the peer review process
altogether in order to promote their own or their associates' work.
Several case reports of dysfunction and breakdown of the peer review
process in the mathematical and physical sciences have recently appeared
in the literature. \cite{Baez 2006,Schiermeier 2008,Trabino 2009}.

Authors, editors and referees are not paid for their work in this
publication process. Nevertheless, publishers often charge exorbitant
amounts for the resulting product, journals which have typically ended
up being hidden away in university libraries, inaccessible to the
public who funded the research in the first place. Independent workers
as well as researchers in poor countries thus have often been cut
out of the research loop entirely.

The need for a system of open electronic publishing of scientific
articles has long been recognized (see e.g. \cite{Odlyzko 1995}).
Several electronic journals have now been created. Some of these charge
readers for access, others are free to read but charge authors for
publication, and still others are free for all parties involved. Perhaps
the biggest success of the Open Access movement was a 2007 U.S. law
requiring all NIH-supported research to be submitted to an openly
accessible archive one year after publication. \cite{Weiss 2007} 

Internet-based alternatives to the prevalent peer review and publishing
process have been discussed in \cite{Harnard 2000} and \cite{Nielsen 2008}.
A trial in open peer review at the journal \emph{Nature} in 2006 generated
widespread debate of the concept \cite{Nature 2006-1}; the final
report concluded that, while the general concept was received enthusiastically,
participation in and satisfaction with their particular model of open
commentary were disappointing. \cite{Nature 2006-2}

\section{ArXiv.org}

The website \noun{arXiv.org} (formerly \noun{xxx.lanl.gov}) is an
electronic archive of freely accessible research preprints. \cite{Ginsparg 1997}
It was started by physicist Paul Ginsparg in August 1991 and has since
become an indispensable tool for researchers in physics, mathematics
and, increasingly, computer science and quantitative biology. Authors
submit their articles to the archive prior to peer review and official
publication by a scientific journal; the preprints are posted on the
website in perpetuity after superficial moderator review. To participate,
authors need an affiliation with a recognized academic institution
or an endorsement by an established author. Interested parties can
sign up for regular e-mail announcements containing the abstracts
of new preprints in their chosen fields.

Once a manuscript has been peer reviewed and accepted for publication,
authors should ideally post an updated version to the archive. Not
all authors remember to do this, and some journals explicitly prohibit
the practice, claiming a copyright on the final result of peer review.%
\footnote{See for instance Elsevier's policy on electronic preprints at  \url{http://www.elsevier.com/wps/find/authorshome.authors/preprints}
(accessed 26 December 2008)%
}

Consequently, \noun{arXiv.org} in its present incarnation and similar
preprint archives in other fields do not serve as authoritative Open
Access repositories of peer reviewed research.

\section{A proposed solution}

To address the problems outlined in section 1, I propose the following
extension to the \noun{arXiv.org} preprint archive. A new class of
users is created, the {}``editors''. Each editor works for an electronic
journal. Authors, after having uploaded a preprint to the archive,
may elect to submit their article for review and official publication
in one of these electronic journals. An editor of that journal then
decides whether the article is appropriate for the journal in terms
of scope and quality. If it is not, this decision is publicly attached
to the article and the process ends; if it is, the editor invites
one or more referees to write public reviews, to be attached to the
article. The article author may subsequently post a public rebuttal
to the reviews. Based on the referee reports and rebuttals, the editor
decides whether to accept, demand changes to, or reject the article.
The original article, reviews, rebuttal and publication decision are
published in perpetuity. If accepted, the author posts a final version
of the article to arXiv.org; as a peer reviewed and officially published
article, it is visibly set apart from mere preprints and added to
the electronic journal's collection of published articles. Rejected
articles may be submitted to another electronic journal.

Reviews should be signed with the referee's full name and affiliation.
This maximizes transparency and allows referees to receive academic
credit for their work. However, some reviewers might be reluctant
to participate in such a system, for instance because they hesitate
to openly reject the work of friends or influential researchers, or
because they do not want to call attention to their ignorance of some
of the issues discussed in the reviewed article. Thus it is probably
necessary to offer referees the option to publish their reviews under
a pseudonym. Over time, such a pseudonym might naturally develop a
reputation as a solid reviewer, completely divorced from the writer's
real-world identity. Using a straightforward cryptographic scheme,
a referee could prove to selected others that he or she owns a certain
pseudonym; in this way even pseudonymous referees could receive academic
credit for their work at the time of tenure or promotion decisions.

Some electronic journals may wish to develop a process for attaching
notes to published articles, for instance to point out prior work,
mistakes or scientific misconduct discovered after publication. It
will also be desirable to attach a moderated discussion forum to each
article, as a natural gathering place of interested researchers. The
quality of these forums would serve as a criterion to differentiate
electronic journals from each other. The pseudonyms used for refereeing
could also be used to sign forum contributions.

One may hope that the proposed system will engender several desirable
consequences. The act of refereeing will rise in prestige in accordance
with its importance for the scientific process. The quality of referee
reports will improve. Outside evaluations and comparisons of the standards
and practices of different electronic journals will become possible.
The process becomes completely transparent and its results are made
freely available.

\end{document}